\documentclass[conference]{IEEEtran}
\IEEEoverridecommandlockouts
\usepackage{cite}
\usepackage{amsmath,amssymb,amsfonts}
\usepackage{algorithmic}
\usepackage{graphicx}
\usepackage{textcomp}
\usepackage{xcolor}
\def\BibTeX{{\rm B\kern-.05em{\sc i\kern-.025em b}\kern-.08em
    T\kern-.1667em\lower.7ex\hbox{E}\kern-.125emX}}
\begin{document}

\title{MicroTEE: Designing TEE OS Based on the Microkernel Architecture\\
\thanks{*Corresponding author: Qianying Zhang, Capital Normal University, No.105, West 3rd Ring North Road, Haidian District, Beijing, China.}
}

\author{
\IEEEauthorblockN{1\textsuperscript{st} Dongxu Ji\quad\quad\quad}
\IEEEauthorblockA{\textit{College of Information Engineering\quad\quad\quad} \\
\textit{Capital Normal University\quad\quad\quad}\\
Beijing, China\quad\quad\quad \\
jidongxu1993@gmail.com\quad\quad\quad}
\and
\IEEEauthorblockN{2\textsuperscript{nd} Qianying Zhang*\quad\quad\quad}
\IEEEauthorblockA{\textit{College of Information Engineering\quad\quad\quad} \\
\textit{Capital Normal University\quad\quad\quad}\\
Beijing, China\quad\quad\quad \\
qyzhang@cnu.edu.cn\quad\quad\quad}
\and
\IEEEauthorblockN{3\textsuperscript{rd} Shijun Zhao\quad\quad\quad}
\IEEEauthorblockA{\textit{Institute of Software\quad\quad\quad} \\
\textit{Chinese Academy of Sciences\quad\quad\quad}\\
Beijing, China\quad\quad\quad \\
zqyzsj@gmail.com\quad\quad\quad}
\and
\IEEEauthorblockN{\quad\quad\quad\quad\quad\quad 4\textsuperscript{th} Zhiping Shi}
\IEEEauthorblockA{\quad\quad\quad\quad\quad\quad \textit{College of Information Engineering} \\
\quad\quad\quad\quad\quad\quad \textit{Capital Normal University}\\
\quad\quad\quad\quad\quad\quad Beijing, China \\
\quad\quad\quad\quad\quad\quad shizp@cnu.edu.cn}
\and
\IEEEauthorblockN{\quad\quad\quad 5\textsuperscript{th} Yong Guan}
\IEEEauthorblockA{\quad\quad\quad \textit{College of Information Engineering} \\
\quad\quad\quad \textit{Capital Normal University}\\
\quad\quad\quad Beijing, China \\
\quad\quad\quad guanyong@cnu.edu.cn}
}

\maketitle

\begin{abstract}
ARM TrustZone technology is widely used to provide Trusted Execution Environments (TEE) for mobile devices. However, most TEE OSes are implemented as monolithic kernels. In such designs, device drivers, kernel services and kernel modules all run in the kernel,  which results in large size of the kernel. It is difficult to guarantee that all components of the kernel have no security vulnerabilities in the monolithic kernel architecture, such as the integer overflow vulnerability in Qualcomm QSEE TrustZone and the TZDriver vulnerability in HUAWEI Hisilicon TEE architecture. This paper presents MicroTEE, a TEE OS based on the microkernel architecture. In MicroTEE, the microkernel provides strong isolation for TEE OS's basic services, such as crypto service and platform key management service. The kernel is only responsible for providing core services such as address space management, thread management, and inter-process communication. Other fundamental services, such as crypto service and platform key management service are implemented as applications at the user layer. Crypto Services and Key Management are used to provide Trusted Applications (TAs) with sensitive information encryption, data signing, and platform attestation functions. Our design avoids the compromise of the whole TEE OS if only one kernel service is vulnerable. A monitor has also been added to perform the switch between the secure world and the normal world. Finally, we implemented a MicroTEE prototype on the Freescale i.MX6Q Sabre Lite development board and tested its performance. Evaluation results show that the performance of cryptographic operations in MicroTEE is better than it in Linux when the size of data is small.
\end{abstract}


\section{Introduction}
Mobile devices have been applied in many fields and can be used to provide online payment, electronic banking and other security-sensitive services. Because mobile devices store a lot of data of great value, it becomes the focus of the attack. Therefore, the security of mobile devices is very important. The ARM TrustZone technology\cite{arm2009security,ngabonziza2016trustzone} is widely used in mobile devices to provide trusted execution environments for security-sensitive services\cite{yalew2017t2droid,hein2015secure,zhang2015trusttokenf,liu2017protc}. And it has become the foundation of the security of mobile devices. Mainstream OEMs like HUAWEI, SAMSUNG and Apple use TrustZone to enhance the security of their products. However, most TEE OSes are implemented based on monolithic kernel architecture. The monolithic kernel architecture is more efficient than microkernel, since components such as drivers and kernel services run in the same address space. However, if any of the kernel services were vulnerable, the whole TEE OS would be compromised. The vulnerability in QSEE \cite{rosenberg2014qsee}, for example, can cause QSEE to write controlled data to arbitrary secure memory. This vulnerability can be exploited to execute arbitrary code in the context of QSEE. By executing malicious code in the secure kernel, key extraction, Linux kernel hijacking and bootLoader unlocking can be achieved. For another example, in HUAWEI Hisilicon's TEE architecture, TrustedCore, a mistake in input structure bound-chek may lead to arbitrary code execution in TEE, while system call bugs in TEE kernel may allow a malicious TA to escalate privilege \cite{2015trustcore}.

Microkernel architecture is proposed to solve the above problems of the monolithic kernel. In the microkernel architecture, the kernel is only responsible for providing core services such as address space management, thread management, and inter-process communication. Other services such as device drivers are implemented as user layer applications. The isolation of the applications is guaranteed by the microkernel. As long as the kernel does not crash, errors in one user layer component will not affect operations of other components. This avoids the compromise of the whole TEE OS if only one kernel service is broken. There are currently some works on microkernels, such as NOVA\cite{steinberg2010nova}, Fiasco.OC\cite{fiasco}, and Redox\cite{redox}.

In this paper, we design a TEE OS based on the microkernel, and then implement the necessary services of a TEE system in the application layer, such as TA management, cryptography services and platform key management. The microkernel provides an isolation mechanism for the above services and TAs at the user layer. The design solves the security issue that exists in the monolithic kernel: the whole TEE OS will compromise if only one kernel service is vulnerable. Our contributions in this paper are as follows.
\begin{itemize}

\item We build a TEE OS, MicroTEE, based on the microkernel architecture. MicroTEE leverages the isolation mechanism provided by the microkernel to isolate basic services of the TEE OS.
\item At the user layer, MicroTEE provides trusted applications with security services such as Crypto Services and Key Management, as well as trusted applications that provide services to the normal world. Crypto services are used to provide TAs with basic cryptographic functions such as encrypting, signing and hashing. Key Management service is used to encrypt sensitive information related to the platform, and it is also used to prove the identity of the device.
\item A monitor is then added to be responsible for the switch between the two worlds, including saving and restoring the context and changing the NS bit.
\item We implement a prototype of MicroTEE on the Freescale i.MX6Q Sabre Lite development board and evaluate its performance.
\end{itemize}

The rest of this paper is organized as follows: Section II describes other works of building a trusted execution environment. Section III introduces the information related to this paper. Section IV describes the design details of the MicroTEE architecture. Section V describes the implementation of MicroTEE. Section VI evaluates the prototype. Finally, we conclude our paper and describe the future work in Section VII.

\section{Related Work}

This section describes other works of building a trusted execution environment. Many studies have focused on providing a trusted execution environment for applications on mobile devices. We first introduce some popular commercial TEE systems, and then introduce some academic and open source TEE systems.
\subsection{Commercial TEE Systems}
 ObC\cite{kostiainen2009board,kostiainen2012board} is a high security, low cost and open certificate management platform designed by Nokia Research Center based on TEE. The software-only management of passwords, keys, certificates and other credential information on mobile devices is not secure enough and the cost of leveraging security hardware to manage credential is too high. ObC can realize the unified management of the whole life cycle\cite{kostiainen2011towards} such as the generation of user credential information, secure storage and secure migration of user credentials information on mobile devices. Based on this platform, ObC has realized applications such as platform attestation \cite{kostiainen2010key,kostiainen2011practical} and public transport ticket system. Kinibi\cite{trustonic} is a TEE implementation built by Trustonic (also known as t-base or Mobicore), and it has been widely used in SamSung and Mediatek devices. T6\cite{trustkernelT6} is a trusted kernel implemented by Shanghai Jiao Tong University based on TrustZone technology. T6 is designed to build an easy-to-use trusted computing platform that provides a high-quality TEE for mobile devices. Qualcomm QSEE is also widely used in mobile devices of various manufacturers, such as SAMSUNG, ASUS and HTC.

\subsection{Academic and Open Source TEE Systems}
TLR\cite{santos2014using,santos2011trusted} is a lightweight, trusted language runtime framework with programming interfaces designed and implemented by Microsoft Research and University of Lisbon for high-level languages such as C\#. TLR allows developers to separate the secure sensitive logic of an application from the rest of the application. The support offered by ARM TrustZone combined with the flexibility of the .NET programming environments allows TLR to offer a secure, yet rich programming environment for developing trusted mobile applications. OP-TEE\cite{linaro2019optee} is an open source project developed by the open source organization Linaro. This project covers all the software components required by TEE software architecture: the user layer client in the normal world, the TEE device driver in the Linux kernel and the trusted OS in the secure world. OP-TEE conforms to the GlobalPlatform specifications and provides standard APIs for the development of TA. ANDIX OS\cite{fitzek2015andix} developed by Graz University of Technology is a multitasking TEE OS. ANDIX OS is compatible with the GlobalPlatform specifications. It realizes the isolation of secure tasks and the mobile operating system through TrustZone technology. Open-TEE\cite{mcgillion2015open}, a virtual, hardware independent TEE implemented in software, is developed by the University of Helsinki. Open-TEE conforms to GlobalPlatform specifications\cite{gpspecification}. It allows developers to develop and debug trusted applications with the same tools they use for developing software. Open-TEE is designed to function as a daemon process in user space, and it mainly includes two components: manager and launcher. Open-TEE ensures that trusted applications developed by it can run on any TEE hardware that conforms to the GlobalPlatform specifications.

From the above works, we can see that most TEE OSes are implemented as monolithic kernels. Although they have relatively high execution efficiency, they cannot avoid the shortage of monolithic kernel which increases attack surfaces due to the large size of the kernel. There is a lack of separation of privileges in the monolithic kernel architecture. A vulnerability of a component in the TEE OS can directly cause the entire system to be taken over by an attacker. For example, the vulnerability found in QSEE can be exploited to execute arbitrary code. We designed the MicroTEE using the memory isolation and protection mechanisms provided by the microkernel architecture. It can implement device drivers and applications at the user layer, using a microkernel to ensure the isolation of the applications. As long as the kernel is not compromised, a problem with one component will not compromise the security of the whole system.
\section{Preliminaries}

\subsection{ARM TrustZone Technology}\label{AA}
ARM TrustZone technology is a set of hardware extensions for ARM processors. It allows the device to run in two different processor modes, called normal world mode and secure world mode. Each mode represents a virtual processor that provides a separate execution environment. Software and hardware resources, such as peripherals and memory, are also divided into two parts. Secure world is usually used to run the security critical software components. It can access all the software and hardware resources, including the resources of the normal world. Normal world can only access the hardware and software resources of the normal world. The NS bit is used to indicate which world the processor is running in. The memory and peripheral controllers use NS bit to check whether a particular resource request is allowed.

TrustZone technology adds a mode to the ARM processor, called monitor mode(Figure 1), which is responsible for the switch between the secure world and the normal world. The monitor mode belongs to the secure world. The entry to monitor can be triggered by executing a dedicated instruction, the Secure Monitor Call (SMC) instruction, or by a subset of the hardware exception mechanisms. The IRQ, FIQ, external Data Abort, and external Prefetch Abort exceptions can all be configured to cause the processor to switch into the monitor mode. The monitor is responsible for saving the context of the current world and switching back to the context of the other world.
\begin{figure}[htbp]
\centerline{\includegraphics[width=9cm]{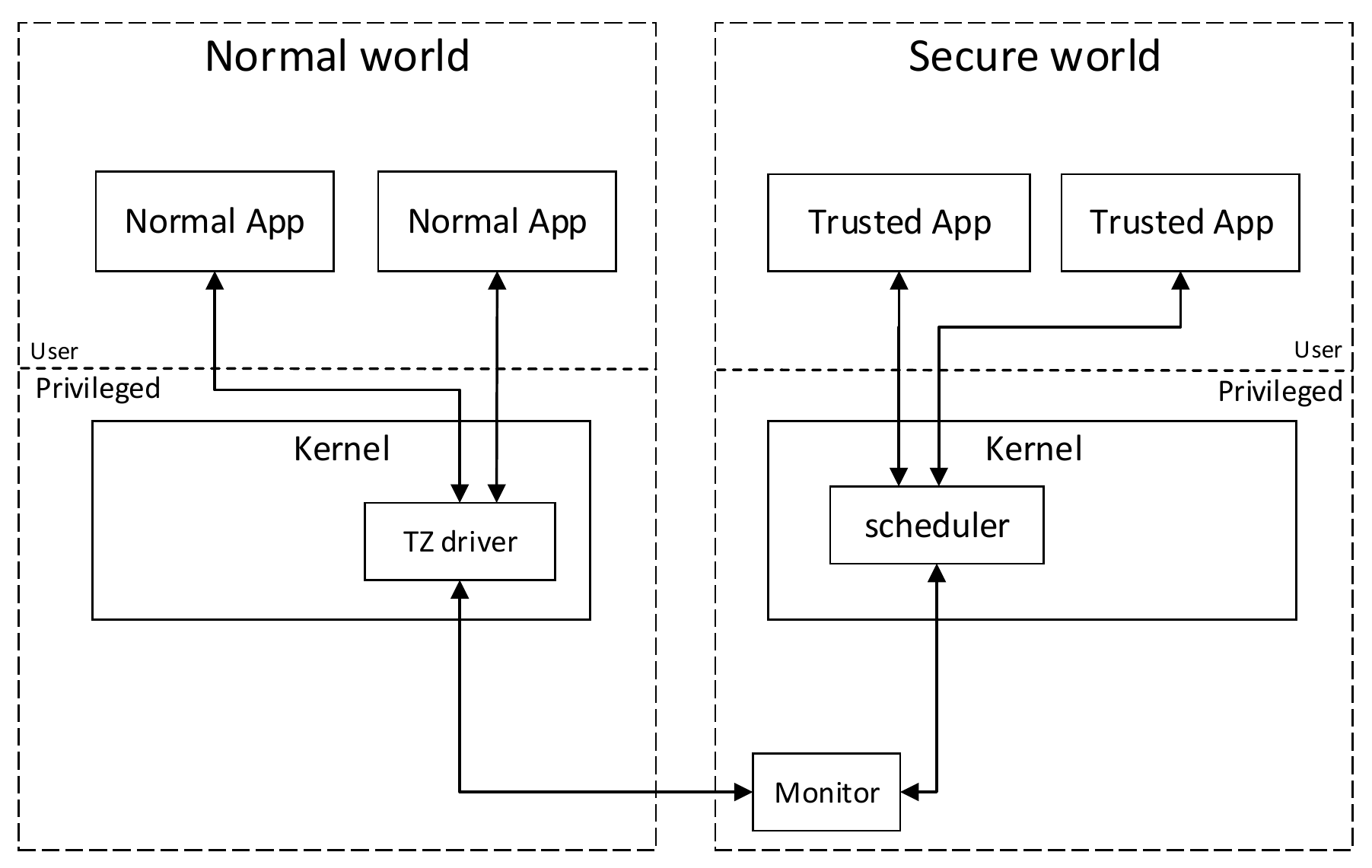}}
\caption{The ARM TrustZone Architecture}
\label{Fig.1}
\end{figure}

Because TrustZone cannot resist physical attacks, studies have provided ways to improve it. The design in\cite{providing2014} leverages SRAM to provide a root of trust for TrustZone. It provides security against both physical and software attacks. SoftME\cite{softme2019}, CaSE\cite{case2016}, TrustShadow\cite{guan2017trustshaow}, and CryptMe\cite{cao2018cryptme} provide TEE system with approaches to resist physical attacks.
\subsection{SEL4 Microkernel}
The seL4\cite{elphinstone2013l3,derrin2006sel4} microkernel is an operating system kernel designed to be a secure, safe, and reliable foundation for systems in a wide variety of application domains. As a microkernel, seL4 provides a small number of services to applications, such as creating and managing virtual address spaces, threads management, and inter-process communication (IPC). The kernel is implemented in approximately 8700 lines of C code. Other services, such as device drivers, run as normal applications in the independent address space at the user layer. They are typically managed by the kernel. The biggest feature of seL4 is that it is the world's first fully formally verified kernel, proven to be bug-free, and can withstand common attacks such as buffer overflow attacks\cite{klein2009sel4,blackham2011timing}.

The seL4 microkernel provides a capability-based access-control model. In order to perform an operation provided by the kernel, the application must invoke the capability it has, which has sufficient rights for the requested service. Endpoint objects are responsible for the IPC communication between threads. The TCB object is used to describe the corresponding thread, and each TCB has a CSpace and a VSpace. CSpace is used to hold all the capabilities of a thread, and VSpace is used to manage the virtual address space. Both of the CSpace and the VSpace can be shared with other threads.

\section{Architecture Design}
This section describes the details of the MicroTEE design, including the design of each component, the relationships between components, the communication between components, and the boot process.

\subsection{MicroTEE Overview}
 The main components of MicroTEE are TEE OS, the Monitor, Root Task and security services. Its architecture is shown in Figure 2. The TEE OS is responsible for providing the fundamental core services, such as TA creation and management, address space management and inter-process communication. The Monitor is responsible for the switch between the secure world and the normal world, including saving and restoring the context of the corresponding world, passing parameters and data, and so on. The Root Task is the first user layer application that runs after kernel startup. It is responsible for taking over all unused resources of the system, creating and managing other TAs. Security services can provide TAs with basic security services such as data encryption and decryption, key management and signing. The components communicate with each other by calling the API of the IPC mechanism provided by the TEE OS. When an interrupt is triggered, the kernel is responsible for sending the interrupt signal to the corresponding handler.
\begin{figure}[htbp]
\centerline{\includegraphics[width=9cm]{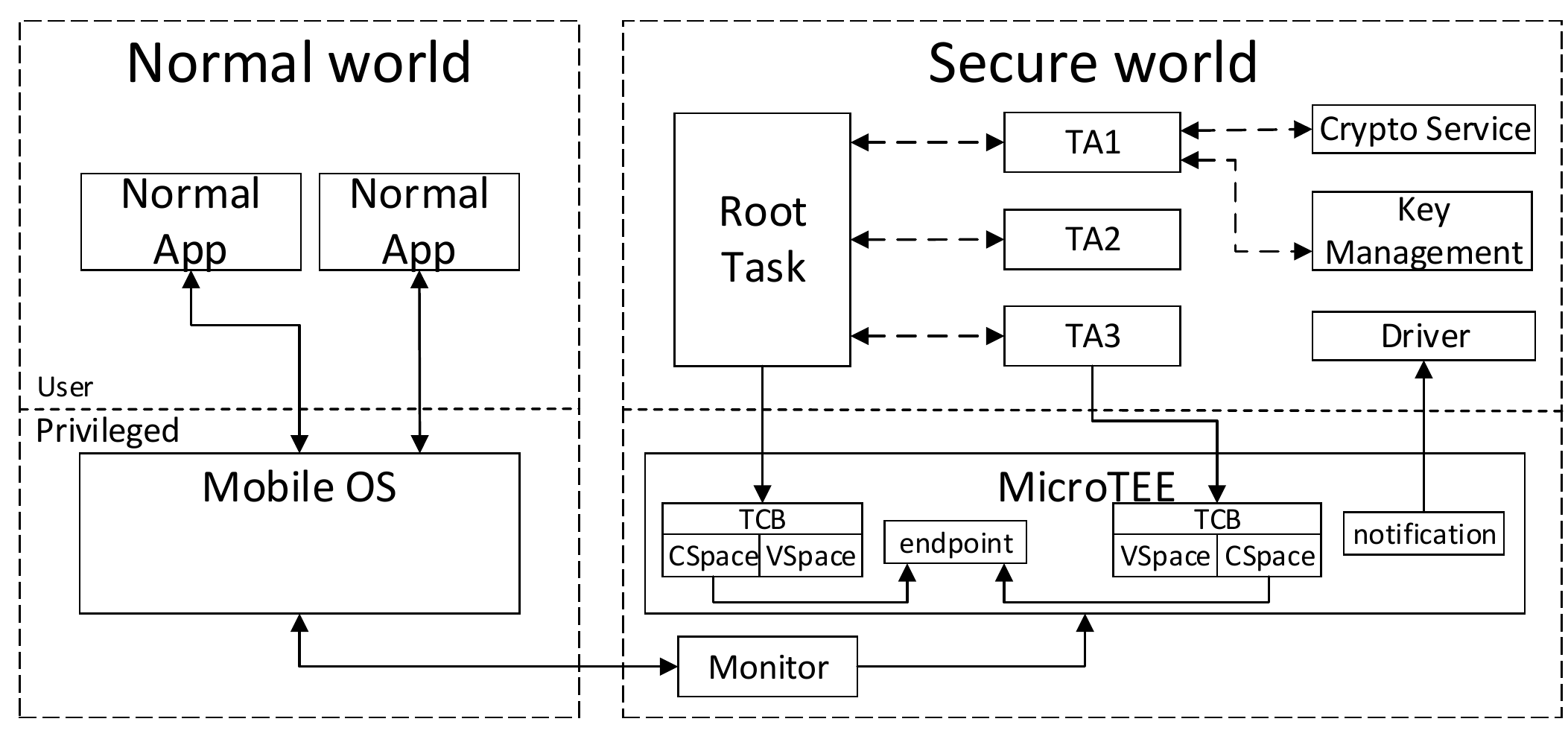}}
\caption{The Architecture of MicroTEE}
\label{Fig.2}
\end{figure}

\subsection{The Monitor}
The monitor is responsible for the switch between the secure world and the normal world. It can be entered into by the SMC instruction. In MicroTEE, when an SMC instruction is invoked, the CPU switches to the monitor mode. Then the corresponding handler, \emph{SMC\_handler}, is invoked according to the monitor vector table base address stored in the Monitor Vector Base Address Register (MVBAR). The current state of the system, whether secure or normal, is depended on the NS bit in the Secure Configuration Register (SCR), and \emph{SMC\_handler} stores the current world's context on the stack. The monitor modifies the NS bit to the corresponding value and loads the context of the corresponding world. We add a new SMC system call, \emph{MicroTEE\_SMC()}, that allows applications at the user layer to enter the privileged layer and then execute the SMC instruction to trigger the world switch. After entering into the monitor mode, the \emph{SMC\_handler} is executed to save and restore the context and pass the command ID and associated parameters.

\subsection{TEE OS}
Our TEE OS is implemented based on the microkernel architecture. As a secure kernel, the microkernel provides a small number of core services for applications, such as address space management, TAs management and inter-process communication.

\textbf{TAs Management.} MicroTEE provides a thread control block (TCB) that manages the information about each TA, including stack pointer, program counter, register values and priority. The kernel guarantees that the information of one TA will not be accessed or changed by other TAs. This enables MicroTEE to provide a separate address space for each TA and allows MicroTEE to ensure that each TA is isolated from each other. MicroTEE schedules each TA according to its priority.

\textbf{Communication Mechanism.} Microkernels typically provide a kernel service for IPC mechanism, such as endpoint object in seL4. In MicroTEE, the IPC communication mechanism is used for communication between TAs and security services. The Root Task creates a kernel service between it and each TA for IPC communication. The Root Task also creates a kernel service for communication between each TA and each security service. TCBs of TAs and security services are configured when the Root Task creates them. The configuration includes the permission to call the kernel service for IPC communication. The sender places the message in a mailbox-like kernel service via a system call, and the receiver receives the corresponding message via another system call. In this way TAs can provide services to the normal world through the Root Task. TAs can also request services from security services.

\textbf{Memory Management.} At boot time, MicroTEE pre-allocates the memory required by the kernel itself, including the code, data, and stack sections. MicroTEE kernel then creates a Root Task and hands over the remaining unused memory to the Root Task. TAs and security services are created and managed by the Root Task. When a new TA needs to be created, the Root Task creates the corresponding TCB first. The page table is then created for the new TA and mapped to the new TA by saving a pointer to the root page directory into the TCB. When a TA is destroyed, MicroTEE destroys the TCB that represents the TA. MicroTEE removes the data from the memory and then retrieves the memory.

\textbf{Interrupt Handling.} In the design of MicroTEE, interrupts are registered in the kernel. When an interrupt is triggered, an interrupt notification signal is sent through the kernel to the particular interrupt handler. And then the handler processes the interrupt. When the user space handler completes interrupt processing, it sends a signal to the kernel, informing that the kernel can send new interrupts to the handler.

\subsection{Root Task}
In MicroTEE, the Root Task is the first user layer application that runs after the kernel boots. It is created and provided with the minimal boot environment by the kernel. This environment consists of the Root Task's TCB, CSpace and VSpace. After the kernel boots, all unused resources are managed by the Root Task, so the Root Task has the highest priority at the user layer. Other TAs are created and managed by the Root Task. Therefore, we treat the Root Task as a manager. When the normal world sends a request, it passes parameters, including the command ID and args, to the secure world through the monitor. The Root Task sends a service request to the appropriate TA according to the command ID received from the normal world. If the command ID does not exist, the Root Task ignores this request. By calling the SMC instruction from the Root Task, the processor switches back to the normal world.

\subsection{Security Services}
MicroTEE provides TA with two security services, Key Management and Crypto Services. They are used to provide key and cryptography services to TA, respectively.

\textbf{Key Management.} This security service provides services related to the platform's keys (assuming that they are provided by the vendor and already stored in trusted hardware), including a Root Key and a Device Key. The root key is a symmetric key used for encryption and decryption of the confidential information, such as keys of TAs. The Device Key is an asymmetric key used to prove the identity of the device, by which a remote entity can verify whether the device is legitimate or not.

\textbf{Crypto Service.} It is responsible for providing basic cryptography service. We provide TAs with cryptographic algorithms by this crypto service, including symmetric encryption algorithm AES (128bits), asymmetric encryption algorithm RSA (2048bits) and hash algorithm SHA-256. A TA requests the corresponding algorithm by sending a request to the Crypto Services through the IPC mechanism. When this service completes the operation, the result is returned to the TA.

\subsection{Trusted Applications}
Each TA is created by the Root Task and has its own address space. When a TA is created, the Root Task assigns a unique command ID to the TA. The functionality of a TA can be customized to provide a service to the normal world. Here, we implement a sample service which increases 1 to the value it receives. When the normal world needs to request this service, it transfers the command ID and parameters to the service by invoking the SMC instruction. The Root Task receives command ID 3 and parameter 0 from the normal world, and then sends a request to the corresponding TA. Finally, the Root Task returns the result to the normal world. Figure 3 shows the process of the Root Task calling a TA.
\begin{enumerate}
    \item The Root Task sends a service request to TA3 based on ID 3.
    \item TA3 receives ID 3 and parameter 0 through the IPC mechanism.
    \item TA3 executes a++ for 0.
    \item TA3 returns the result 1 to the Root Task.
    \item The Root Task receives the result.
\end{enumerate}

\begin{figure}[htbp]
\centerline{\includegraphics[width=9cm]{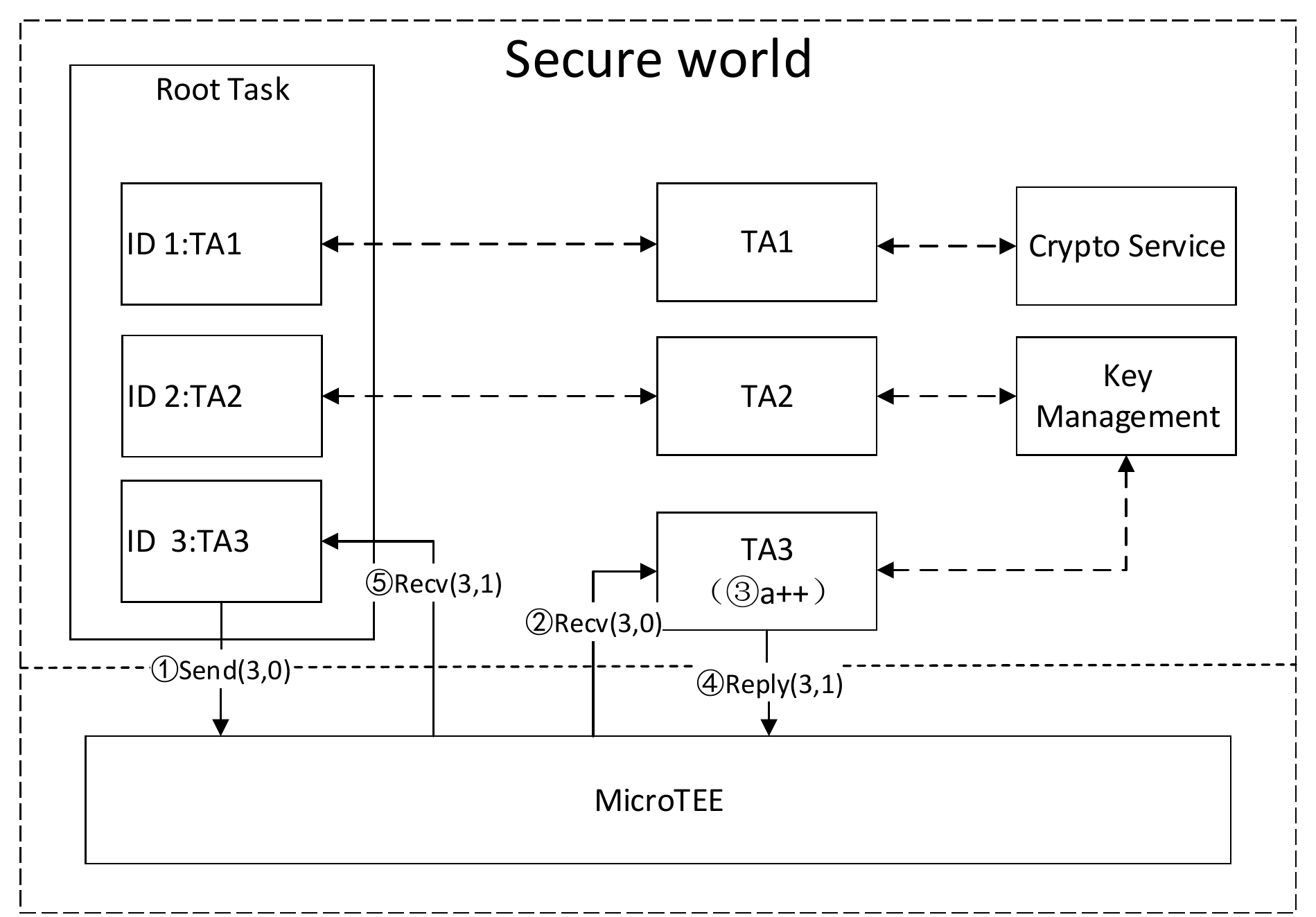}}
\caption{The Process of Root Task Calling TA}
\label{Fig.3}
\end{figure}

\subsection{Secure Boot}
At boot time, attackers can perform downgrade attacks, which replace secure software images in flash with tampered images. If the system boots an image from flash without checking its integrity, the system is vulnerable. So we build a chain of trust for the software in the secure world, which is built from the ROM that is not easily tampered with.

In our design, the boot process is shown in Figure 4. A device starts from the Boot ROM, which loads the first bootloader image, with the manufacturer's public key and a signature at the end of the image. The hash of the public key is hard-coded into the ROM. The ROM first verifies the public key with the hash and then uses the public key to verify the signature of the first bootloader image. After the verification succeeds, the first bootloader loads the secure world image and second bootloader image into memory and verifies the integrity of the images. If the verification succeeds, the bootloader transfers its control to the TEE OS. After that, the second bootloader loads the mobile OS image and verifies it. The secure boot guarantees that the initial state of the device is trusted.
\begin{figure}[htbp]
\centerline{\includegraphics[width=9cm]{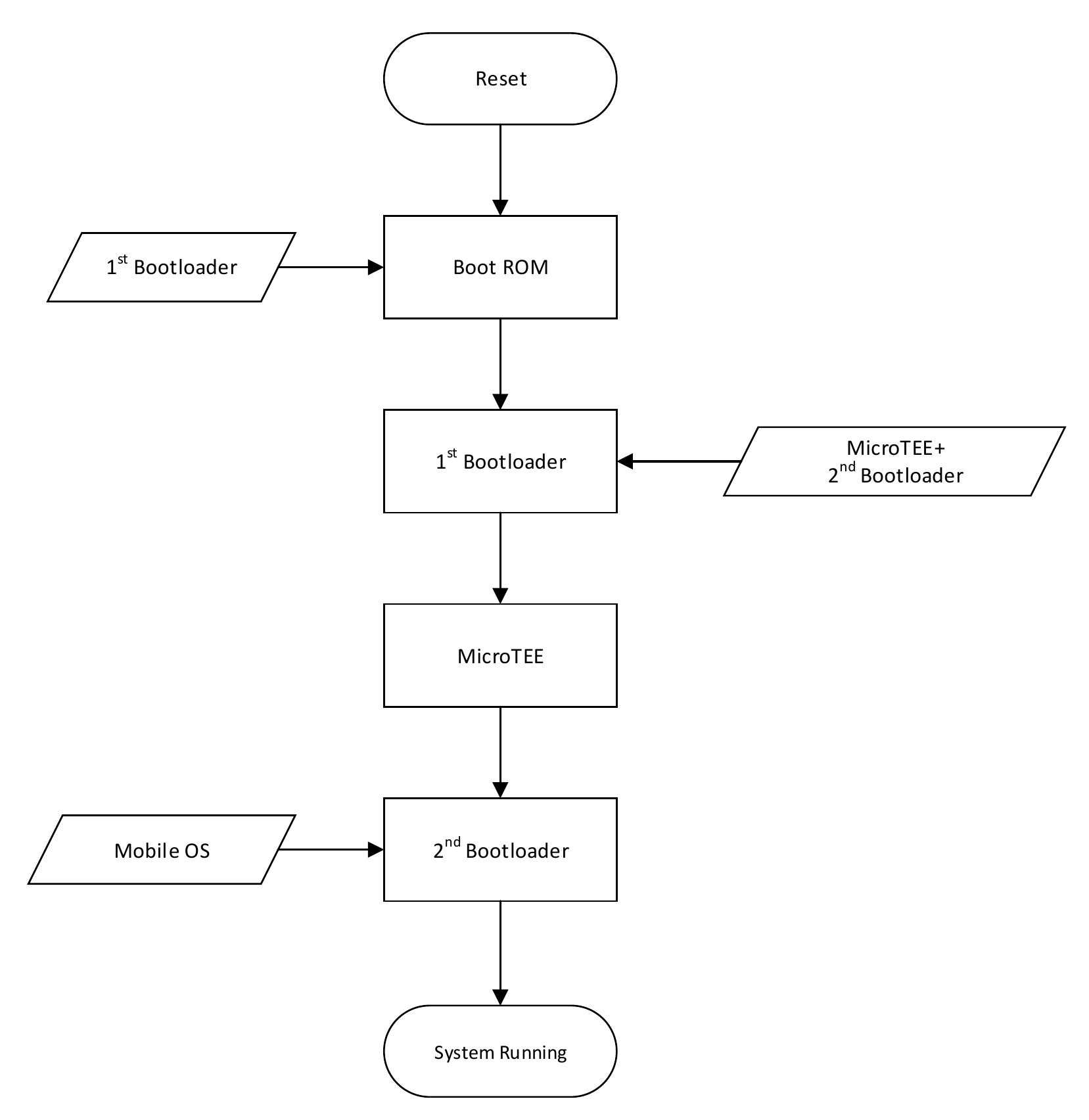}}
\caption{Secure Boot Process}
\label{Fig.4}
\end{figure}

\section{Prototype Implementation}
We have implemented a MicroTEE prototype on the Freescale i.MX6Q Sabre Lite development board. This section describes the implementation of all the components, including the Monitor, the MicroTEE kernel based on the seL4 microkernel and the security services.

\subsection{The Monitor}
The monitor is responsible for the switch between the normal world and the secure world. It is configured during platform initialization, including setting the stack of monitor mode and writing vector base address into Monitor Vector Base Address Register (MVBAR). The stack is used to store CPU contexts of the normal world and the secure world, and the MVBAR holds separate exceptions table of the monitor mode.

The secure world switches to the monitor mode by invoking the SMC instruction. At the user layer, the processor switches to the privilege mode by calling our newly added system call, \emph{MicroTEE\_SMC(command\_ID, args)}. Then the processor executes the SMC instruction in the privilege mode (SMC can only be invoked in the privilege mode). The parameter \emph{args} is used to pass other information, such as the address of shared memory. The values of these parameters are passed to the TEE OS through the physical registers r0 and r1, because these registers are not changed in the monitor. Finally, the secure world can perform services based on \emph{command\_ID} and data passed from the normal world. The SMC exception handler in the monitor mode, \emph{SMC\_handler()}, does the work of saving the current world CPU context and restoring the other world CPU context. Since only the secure world is studied in this paper, the context of the secure world is saved and restored in the monitor.

\subsection{TEE OS and Security Services}
We build the MicroTEE Kernel based on the seL4 microkernel. MicroTEE provides user layer TAs with TCB to manage the relevant configuration information. It also provides assurance that each TA's address space is isolated from each other. Applications implement IPC communication between them by calling system calls provided by MicroTEE.

In MicroTEE, security services are implemented as user layer applications, including Key Management and Crypto Services. We add LibTomCrypt to our MicroTEE which provides the following cryptographic services: symmetric cryptography algorithm AES, asymmetric cryptography algorithm RSA and hash algorithm SHA-256. A TA sends the request including the service ID and data to the corresponding security service. And then the operation result is returned to the TA. Key Management service manages the platform's Root Key and Device Key, and these Keys don't leave Key Management. Key Management provides encryption of confidential information related to the device and attestation of platform identity. When a TA needs to encrypt confidential information or to prove the identity of the device, the TA sends a request to the Key Management service. The Key Management service returns the result to the TA after the encryption or signature operation is completed. The Crypto Services provides basic cryptography services such as AES, RSA and SHA-256. Similarly, the corresponding algorithm is called according to the request, and the result is returned to the TA.

\subsection{Secure Boot}
After the power of the system is reset, the Boot ROM loads the first bootloader's image into the memory, which includes the first bootloader's native image and digital signature. After the image is loaded into memory, its integrity needs to be verified by Boot ROM. If the verification succeeds, the images of MicroTEE OS and the second bootloader will be loaded into memory by the first bootloader and similar verification steps will be carried out. After the verification, MicroTEE OS will boot. In the boot procedure, the microkernel has to perform an integrity check of the Root Task before launching it. And then the image of mobile OS is loaded into memory and verified by the second bootloader. Finally, mobile OS boots.
\section{Evaluation}

\subsection{Performance of IPC Communication}
The IPC mechanism is only used to send critical short messages. For long messages, a shared memory should be used. We leverage the sel4bench tool to evaluate the performance of IPC. The sel4bench evaluates the performance of IPC when the message length is 0 and 10 words. A TA, as the sender, requests services from the Security Services. The test results are shown in Figure 5, each of which is an average of 16 calls. When the message length is 0, the time consumption for a TA to send the message is 295.8 ns, and the time consumption for the Security Services to return the result is 307.6 ns. When the message length is 10 message words, the time cost for TA to send the message is 862.1 ns, and the time cost for the Security Service to return the result is 853.8 ns. The increase in time cost is caused by reading and writing messages. When the length of the IPC communication message is less than 10, the IPC communication time cost is less than 1 $\mu$s. Therefore, when messages sent through IPC communication are small, the impact of IPC communication on system performance is little.
\begin{figure}[htbp]
\centerline{\includegraphics[width=8.5cm]{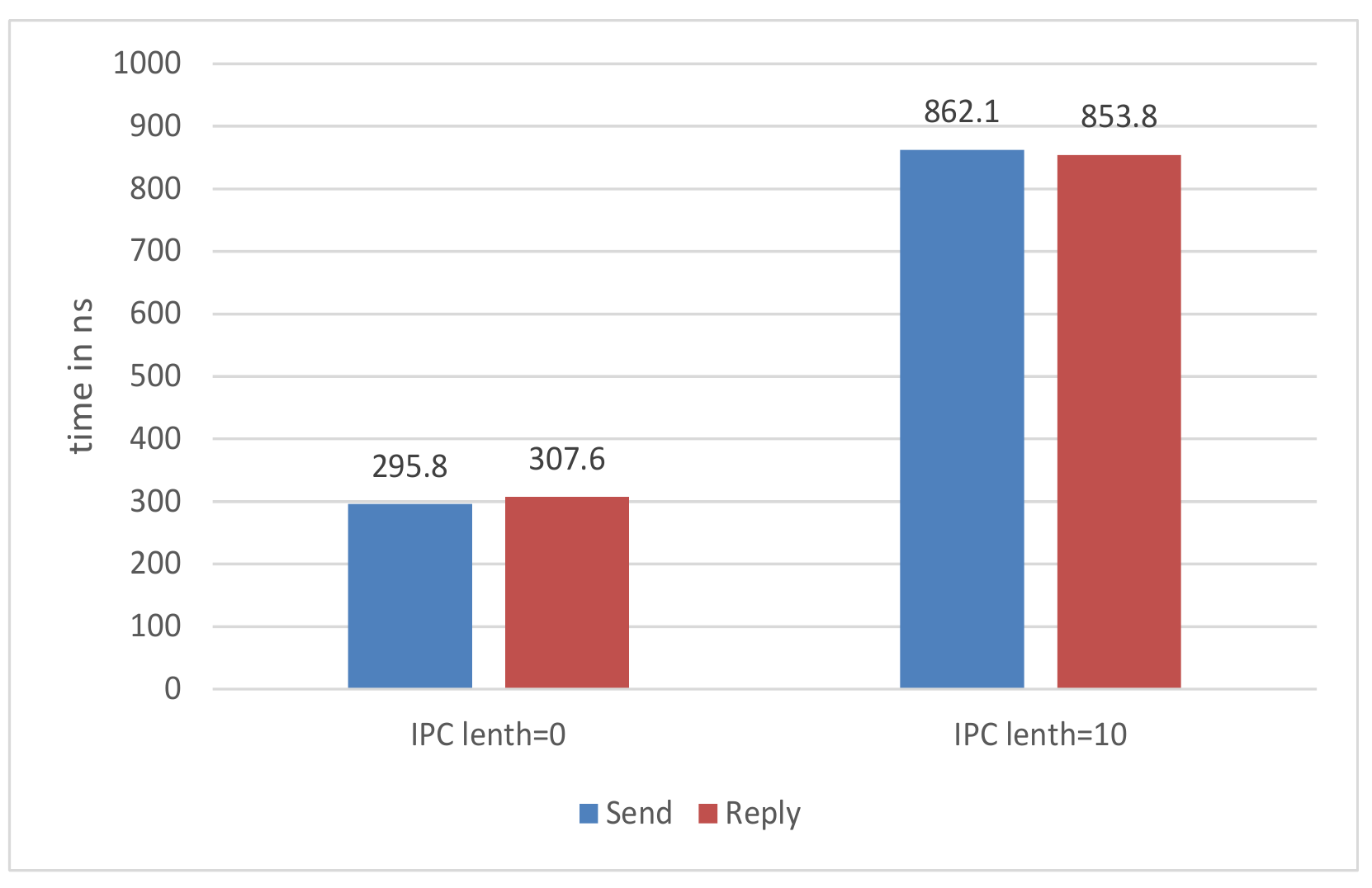}}
\caption{Time of IPC }
\label{Fig.5}
\end{figure}

\subsection{Overhead of World Switch}
 The time cost of world switch is also measured. Measurement begins with a TA executing an SMC system call, \emph{MicroTEE\_SMC(command\_ID, args)}, and ends with the TA continuing to execute. The world switch involves switching into the monitor from the TA and returning to the TA from the monitor. Table 1 shows the overhead of a world switch. Test results show that the cost of the switch between worlds is high. This is because a world switch requires multiple context switches. A context switch from the user layer to the kernel is required, and a context switch from the kernel to the monitor is also required.
\begin{table}[htbp]
\label{tab1}
\caption{SMC Call Overhead}
\begin{center}
\begin{tabular}{|c|c|}
\hline
\textbf{World Switch}&\textbf{Overhead(ms)} \\
\cline{1-2}
\textbf{SMC Call and Return} & \textbf{2.002} \\
\hline
\end{tabular}
\end{center}
\end{table}

\subsection{Cryptographic Services}
We have measured the performance of symmetric cryptography algorithm AES, asymmetric cryptography algorithm RSA and hash algorithm SHA-256 respectively. Figure 6 shows the performance results of symmetric cryptography algorithm AES operating on the data of different sizes in MicroTEE. We can see that the performance of encryption and decryption is similar and stable. Figure 7 shows the performance of AES in Linux. When the size of data is small, the performance of AES in MicroTEE is better than in Linux. The reason is that when measuring time consumption, the system call to get the time introduces a context switch. The context switch of the Linux kernel service takes a large proportion when the size of data is small. The context switch of the microkernel has a fixed overload and is lighter than the context switch of Linux. When the size of data is close to 1MB, cryptographic computations take the most proportion, and the time used for context switch is negligible. So the performance of AES in MicroTEE and Linux is similar.

\begin{figure}[htbp]
\centerline{\includegraphics[width=8.5cm]{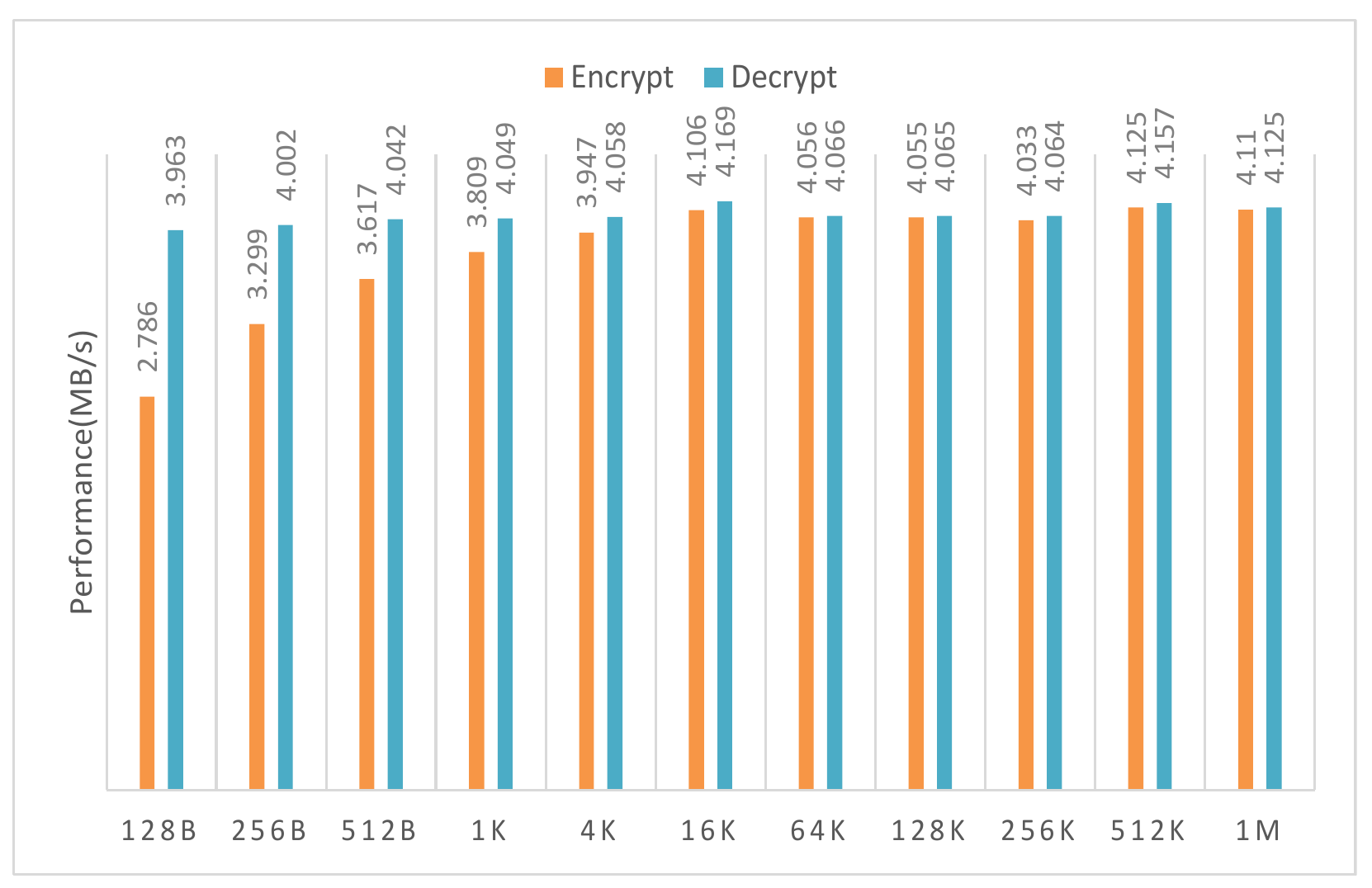}}
\caption{Performance of AES Encryption and Decryption in MicroTEE}
\label{Fig.6}
\end{figure}
\begin{figure}[htbp]
\centerline{\includegraphics[width=8.5cm]{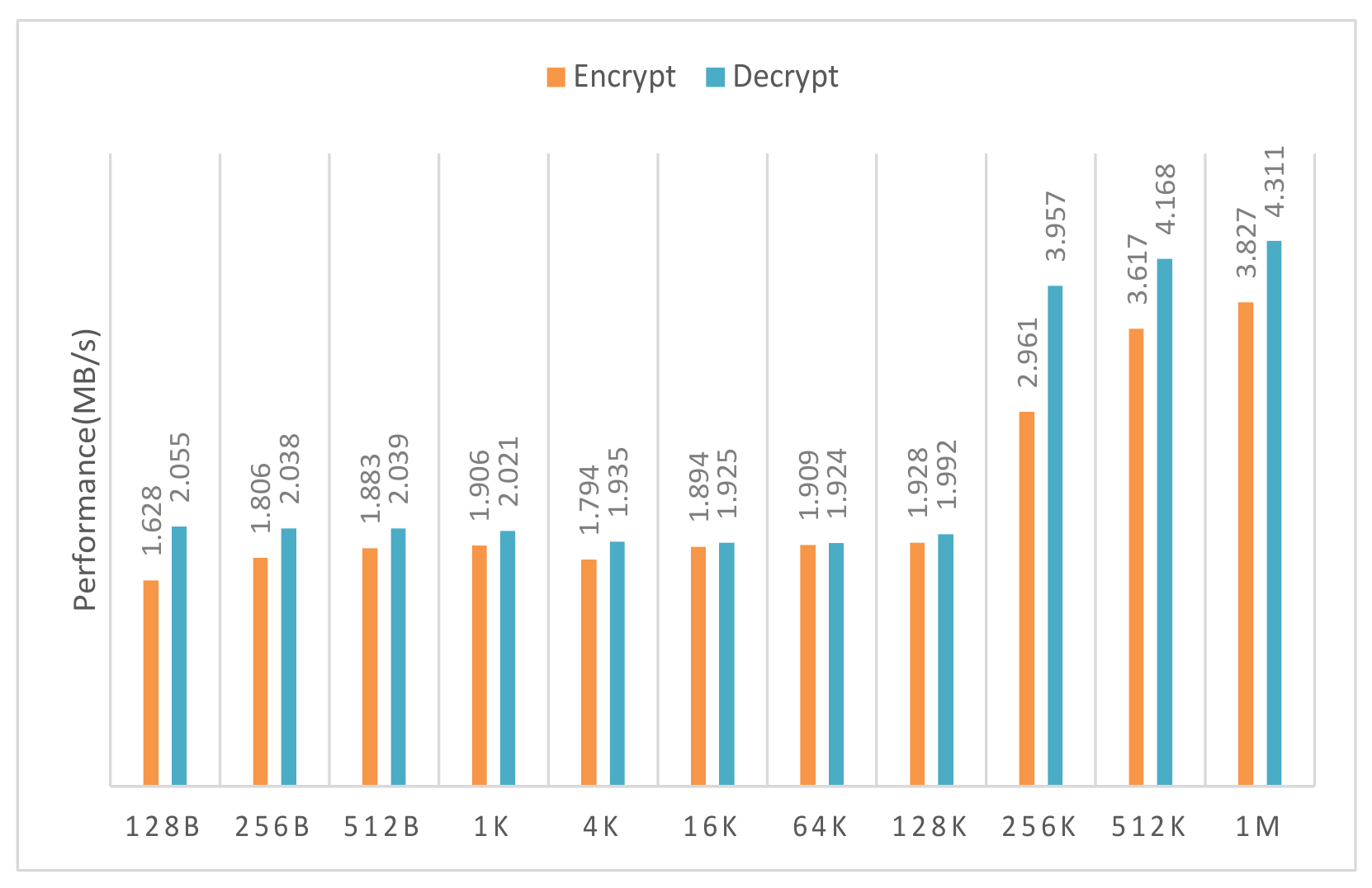}}
\caption{Performance of AES Encryption and Decryption in Linux}
\label{Fig.7}
\end{figure}
Table 2 shows the performance of the signing and verifying operations of the RSA algorithm in MicroTEE, where the length of the key  is 1024 bits and 2048 bits, respectively. When the length of the key is increased from 1024 bits to 2048 bits, the time required for the signing operation is increased by about 4.6 times, and the time required for the verifying operation is increased by about 3.2 times. The choice of key length is the result of a balance of security and performance.
\begin{table}[htbp]
\label{tab2}
\caption{1024\&2048bits-RSA}
\begin{center}
\begin{tabular}{|c|c|c|}
\hline
\textbf{Operation type} & \textbf{1024bits($\mu$s)} &\textbf{2048bits($\mu$s)}\\
\cline{1-3}
\textbf{Sign} & \textbf{13851.4} & \textbf{63821.6}\\
\cline{1-3}
\textbf{Verify} & \textbf{777.6} & \textbf{2482.8}\\
\hline
\end{tabular}
\end{center}
\end{table}

Figure 8 shows the performance of the hash algorithm SHA-256 for the data of different sizes in MicroTEE and Linux. The evaluation result of SHA-256 in MicroTEE and Linux is similar to AES. MicroTEE performs better than Linux when the data size is small. When the data size increases, the time is mainly used for cryptographic calculations. So when the data size is larger than 256K, SHA-256 performs similarly in MicroTEE and Linux. Combined with the performance of AES and SHA-256, MicroTEE has an advantage: the kernel context switch is lightweight and has less impact on application operations.
\begin{figure}[htbp]
\centerline{\includegraphics[width=8.5cm]{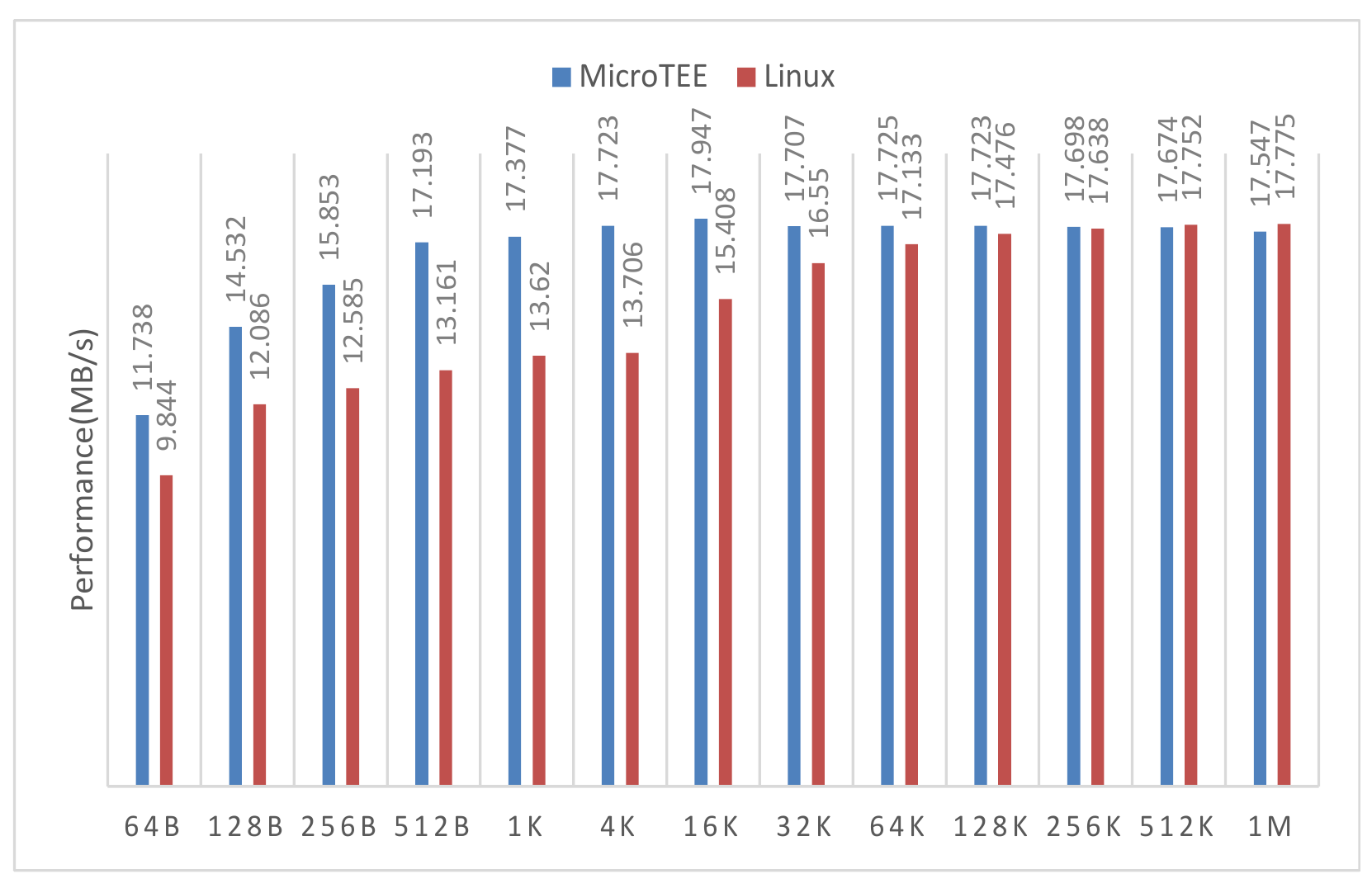}}
\caption{Performance of SHA-256}
\label{Fig.8}
\end{figure}

\section{Conclusions and Future Work}
This paper introduces the design, implementation and evaluation of MicroTEE based on microkernel architecture. In our design, the microkernel is used as the secure kernel to provide the core services for TEE, and other security services necessary for TEE are implemented as user applications of the microkernel. These security services provide TAs with basic cryptography services such as encryption, signing and hash. We have implemented a MicroTEE prototype on the Freescale i.MX6Q Sabre Lite development board. The evaluation shows that the performance in the MicroTEE is stable and it is better than the performance in Linux in the aspect of providing cryptography services when the size of data is small.

In the future, we intend to extend the trusted computing functions on this architecture, including integrity measurement, data sealing and remote attestation to improve the security of our system. Integrity measurement is responsible for verifying the integrity of TA images loaded into memory before execution. Data sealing is responsible for binding information to a particular service. Remote attestation is responsible for proving the state of the device to a remote party.

\section*{Acknowledgment}
This work was supported by the National Natural Science Foundation of China (61602325, 61802375, 61876111, 61877040, 61572331), and the Project of Beijing Municipal Education Commission (KM20190028005).

\bibliographystyle{IEEEtran}
\bibliography{ref}

\begin{thebibliography}{10}
\providecommand{\url}[1]{#1}
\csname url@samestyle\endcsname
\providecommand{\newblock}{\relax}
\providecommand{\bibinfo}[2]{#2}
\providecommand{\BIBentrySTDinterwordspacing}{\spaceskip=0pt\relax}
\providecommand{\BIBentryALTinterwordstretchfactor}{4}
\providecommand{\BIBentryALTinterwordspacing}{\spaceskip=\fontdimen2\font plus
\BIBentryALTinterwordstretchfactor\fontdimen3\font minus
  \fontdimen4\font\relax}
\providecommand{\BIBforeignlanguage}[2]{{%
\expandafter\ifx\csname l@#1\endcsname\relax
\typeout{** WARNING: IEEEtran.bst: No hyphenation pattern has been}%
\typeout{** loaded for the language `#1'. Using the pattern for}%
\typeout{** the default language instead.}%
\else
\language=\csname l@#1\endcsname
\fi
#2}}
\providecommand{\BIBdecl}{\relax}
\BIBdecl

\bibitem{arm2009security}
ARM, ``Security technology building a secure system using trustzone technology
  (white paper),'' \emph{ARM Limited}, 2009.

\bibitem{ngabonziza2016trustzone}
B.~Ngabonziza, D.~Martin, A.~Bailey, H.~Cho, and S.~Martin, ``Trustzone
  explained: Architectural features and use cases,'' in \emph{2016 IEEE 2nd
  International Conference on Collaboration and Internet Computing
  (CIC)}.\hskip 1em plus 0.5em minus 0.4em\relax IEEE, 2016, pp. 445--451.

\bibitem{yalew2017t2droid}
S.~D. Yalew, G.~Q. Maguire, S.~Haridi, and M.~Correia, ``T2droid: A
  trustzone-based dynamic analyser for android applications,'' in \emph{2017
  IEEE Trustcom/BigDataSE/ICESS}.\hskip 1em plus 0.5em minus 0.4em\relax IEEE,
  2017, pp. 240--247.

\bibitem{hein2015secure}
D.~Hein, J.~Winter, and A.~Fitzek, ``Secure block device--secure, flexible, and
  efficient data storage for arm trustzone systems,'' in \emph{2015 IEEE
  Trustcom/BigDataSE/ISPA}, vol.~1.\hskip 1em plus 0.5em minus 0.4em\relax
  IEEE, 2015, pp. 222--229.

\bibitem{zhang2015trusttokenf}
Y.~Zhang, S.~Zhao, Y.~Qin, B.~Yang, and D.~Feng, ``Trusttokenf: A generic
  security framework for mobile two-factor authentication using trustzone,'' in
  \emph{2015 IEEE Trustcom/BigDataSE/ISPA}, vol.~1.\hskip 1em plus 0.5em minus
  0.4em\relax IEEE, 2015, pp. 41--48.

\bibitem{liu2017protc}
R.~Liu and M.~Srivastava, ``Protc: Protecting drone's peripherals through arm
  trustzone,'' in \emph{Proceedings of the 3rd Workshop on Micro Aerial Vehicle
  Networks, Systems, and Applications}.\hskip 1em plus 0.5em minus 0.4em\relax
  ACM, 2017, pp. 1--6.

\bibitem{rosenberg2014qsee}
D.~Rosenberg, ``Qsee trustzone kernel integer over flow vulnerability,'' in
  \emph{Black Hat conference}, 2014, p.~26.

\bibitem{2015trustcore}
D.~Shen, ``Attcaking your "trusted core" exploiting trustzone on android,'' in
  \emph{Black Hat conference}, 2015.

\bibitem{steinberg2010nova}
U.~Steinberg and B.~Kauer, ``Nova: a microhypervisor-based secure
  virtualization architecture,'' in \emph{Proceedings of the 5th European
  conference on Computer systems}.\hskip 1em plus 0.5em minus 0.4em\relax ACM,
  2010, pp. 209--222.

\bibitem{fiasco}
TU-Dresden, ``Fiasco.oc,'' \url{https://os.inf.tu-dresden.de/fiasco/}.

\bibitem{redox}
R.~Developers, ``Redox,'' \url{https://www.redox-os.org/}.

\bibitem{kostiainen2009board}
K.~Kostiainen, J.-E. Ekberg, N.~Asokan, and A.~Rantala, ``On-board credentials
  with open provisioning,'' in \emph{Proceedings of the 4th International
  Symposium on Information, Computer, and Communications Security}.\hskip 1em
  plus 0.5em minus 0.4em\relax ACM, 2009, pp. 104--115.

\bibitem{kostiainen2012board}
K.~Kostiainen \emph{et~al.}, ``On-board credentials: an open credential
  platform for mobile devices,'' 2012.

\bibitem{kostiainen2011towards}
K.~Kostiainen, N.~Asokan, and A.~Afanasyeva, ``Towards user-friendly credential
  transfer on open credential platforms,'' in \emph{International conference on
  Applied cryptography and network security}.\hskip 1em plus 0.5em minus
  0.4em\relax Springer, 2011, pp. 395--412.

\bibitem{kostiainen2010key}
K.~Kostiainen, A.~Dmitrienko, J.-E. Ekberg, A.-R. Sadeghi, and N.~Asokan, ``Key
  attestation from trusted execution environments,'' in \emph{International
  Conference on Trust and Trustworthy Computing}.\hskip 1em plus 0.5em minus
  0.4em\relax Springer, 2010, pp. 30--46.

\bibitem{kostiainen2011practical}
K.~Kostiainen, N.~Asokan, and J.-E. Ekberg, ``Practical property-based
  attestation on mobile devices,'' in \emph{International Conference on Trust
  and Trustworthy Computing}.\hskip 1em plus 0.5em minus 0.4em\relax Springer,
  2011, pp. 78--92.

\bibitem{trustonic}
Trustonic, ``Kinibi,'' \url{https://www.trustonic.com/}.

\bibitem{trustkernelT6}
TurstKernel, ``T6,'' \url{https://www.trustkernel.com/}.

\bibitem{santos2014using}
N.~Santos, H.~Raj, S.~Saroiu, and A.~Wolman, ``Using arm trustzone to build a
  trusted language runtime for mobile applications,'' in \emph{ACM SIGARCH
  Computer Architecture News}, vol.~42, no.~1.\hskip 1em plus 0.5em minus
  0.4em\relax ACM, 2014, pp. 67--80.

\bibitem{santos2011trusted}
N.~Santos, H.~Raj, S.~Saroiu, and Wolman, ``Trusted language runtime (tlr):
  enabling trusted applications on smartphones,'' in \emph{Proceedings of the
  12th Workshop on Mobile Computing Systems and Applications}.\hskip 1em plus
  0.5em minus 0.4em\relax ACM, 2011, pp. 21--26.

\bibitem{linaro2019optee}
Linaro, ``Op-tee documentation,'' 2019.

\bibitem{fitzek2015andix}
A.~Fitzek, F.~Achleitner, J.~Winter, and D.~Hein, ``The andix research os¡ªarm
  trustzone meets industrial control systems security,'' in \emph{2015 IEEE
  13th International Conference on Industrial Informatics (INDIN)}.\hskip 1em
  plus 0.5em minus 0.4em\relax IEEE, 2015, pp. 88--93.

\bibitem{mcgillion2015open}
B.~McGillion, T.~Dettenborn, T.~Nyman, and N.~Asokan, ``Open-tee--an open
  virtual trusted execution environment,'' in \emph{2015 IEEE
  Trustcom/BigDataSE/ISPA}, vol.~1.\hskip 1em plus 0.5em minus 0.4em\relax
  IEEE, 2015, pp. 400--407.

\bibitem{gpspecification}
GlobalPlatform, ``Tee internal core api specification,'' 2018.

\bibitem{providing2014}
S.~Zhao, Q.~Zhang, G.~Hu, Y.~Qin, and D.~Feng, ``Providing root of trust for
  arm trustzone using on-chip sram,'' in \emph{Proceedings of the 4th
  International Workshop on Trustworthy Embedded Devices}.\hskip 1em plus 0.5em
  minus 0.4em\relax ACM, 2014, pp. 25--36.

\bibitem{softme2019}
M.~Zhang, Q.~Zhang, S.~Zhao, Z.~Shi, and Y.~Guan, ``Softme: A software-based
  memory protection approach for tee system to resist physical attacks,''
  \emph{Security and Communication Networks}, 2019.

\bibitem{case2016}
N.~Zhang, K.~Sun, W.~Lou, and Y.~T. Hou, ``Case: Cache-assisted secure
  execution on arm processors,'' in \emph{2016 IEEE Symposium on Security and
  Privacy (SP)}.\hskip 1em plus 0.5em minus 0.4em\relax IEEE, 2016, pp. 72--90.

\bibitem{guan2017trustshaow}
L.~Guan, P.~Liu, X.~Xing, X.~Ge, S.~Zhang, M.~Yu, and T.~Jaeger, ``Trustshadow:
  Secure execution of unmodified applications with arm trustzone,'' in
  \emph{Proceedings of the 15th Annual International Conference on Mobile
  Systems, Applications, and Services}.\hskip 1em plus 0.5em minus 0.4em\relax
  ACM, 2017, pp. 488--501.

\bibitem{cao2018cryptme}
C.~Cao, L.~Guan, N.~Zhang, N.~Gao, J.~Lin, B.~Luo, P.~Liu, J.~Xiang, and
  W.~Lou, ``Cryptme: Data leakage prevention for unmodified programs on arm
  devices,'' in \emph{International Symposium on Research in Attacks,
  Intrusions, and Defenses}.\hskip 1em plus 0.5em minus 0.4em\relax Springer,
  2018, pp. 380--400.

\bibitem{elphinstone2013l3}
K.~Elphinstone and G.~Heiser, ``From l3 to sel4 what have we learnt in 20 years
  of l4 microkernels?'' in \emph{Proceedings of the Twenty-Fourth ACM Symposium
  on Operating Systems Principles}.\hskip 1em plus 0.5em minus 0.4em\relax ACM,
  2013, pp. 133--150.

\bibitem{derrin2006sel4}
P.~Derrin, D.~Elkaduwe, and K.~Elphinstone, ``sel4 reference manual,''
  \emph{NICTA-National Information and Communications Technology Australia},
  2006.

\bibitem{klein2009sel4}
G.~Klein, K.~Elphinstone, G.~Heiser, J.~Andronick, D.~Cock, P.~Derrin,
  D.~Elkaduwe, K.~Engelhardt, R.~Kolanski, M.~Norrish \emph{et~al.}, ``sel4:
  Formal verification of an os kernel,'' in \emph{Proceedings of the ACM SIGOPS
  22nd symposium on Operating systems principles}.\hskip 1em plus 0.5em minus
  0.4em\relax ACM, 2009, pp. 207--220.

\bibitem{blackham2011timing}
B.~Blackham, Y.~Shi, S.~Chattopadhyay, A.~Roychoudhury, and G.~Heiser, ``Timing
  analysis of a protected operating system kernel,'' in \emph{2011 IEEE 32nd
  Real-Time Systems Symposium}.\hskip 1em plus 0.5em minus 0.4em\relax IEEE,
  2011, pp. 339--348.

\end{thebibliography}
\end{document}